\documentclass[journal=jacsat,manuscript=article]{achemso}

\usepackage[version=3]{mhchem} % Formula subscripts using \ce{}
\usepackage[T1]{fontenc}       % Use modern font encodings
\usepackage{dcolumn}

\author{Rafik Smaali}
\affiliation {Clermont Universit\'e, Universit\'e Blaise Pascal, Institut Pascal, BP 10448, F-63000 Clermont-Ferrand, France}
\alsoaffiliation{CNRS, UMR 6602, Institut Pascal, F-63177 Aubi\`ere, France}
\author{Fatima Omeis}
\affiliation {Clermont Universit\'e, Universit\'e Blaise Pascal, Institut Pascal, BP 10448, F-63000 Clermont-Ferrand, France}
\alsoaffiliation{CNRS, UMR 6602, Institut Pascal, F-63177 Aubi\`ere, France}
\author{Antoine Moreau}
\affiliation {Clermont Universit\'e, Universit\'e Blaise Pascal, Institut Pascal, BP 10448, F-63000 Clermont-Ferrand, France}
\alsoaffiliation{CNRS, UMR 6602, Institut Pascal, F-63177 Aubi\`ere, France}
\author{Thierry Taliercio}
\affiliation {Universit\'e Montpellier, IES, UMR 5214, F-34000, Montpellier, France}
\alsoaffiliation {CNRS, IES, UMR 5214, F-34000, Montpellier, France}

\author{Emmanuel Centeno}
\email{emmanuel.centeno@univ-bpclermont.fr}
\affiliation {Clermont Universit\'e, Universit\'e Blaise Pascal, Institut Pascal, BP 10448, F-63000 Clermont-Ferrand, France}
\alsoaffiliation {CNRS, UMR 6602, Institut Pascal, F-63177 Aubi\`ere, France}

\title {Universal metamaterial absorber}


\begin{document}

\begin{abstract}
We propose a design for an universal absorber, characterized by a resonance frequency that can be tuned from visible to microwave frequencies independently of the choice of the metal and the dielectrics involved.  An almost resonant perfect absorption up to $99.8 \%$ is demonstrated at resonance for all polarization states of light and for a very wide angular aperture. These properties originate from a magnetic Fabry-Perot mode that is confined in a dielectric spacer of $\lambda/100$ thickness by a metamaterial layer and a mirror. An extraordinary large funneling through nano-slits explains how light can be trapped in the structure. Simple scaling laws can be used as a recipe to design ultra-thin perfect absorbers whatever the materials and the desired resonance wavelength, making our design truly universal.
\end{abstract}

\section{Introduction}
The control of light absorbance plays a fundamental role in today's photonics technologies with strong impacts for solar energy harvesting or for light emitting and sensing components \cite{Collin:2014hd, Liu:2010kw, Park:2015gm}. Since according the Kirchhoff's law, perfect absorbers and emitters are equivalent, significant efforts are pursued to realize compact artificial materials presenting an almost perfect absorption in a selective spectral range, for any polarization or incidence angle\cite{Shi:2014jb, Hedayati:2014gf}. The targeted operating frequency usually imposes the choice of the materials constituting the absorbers and also strongly constraints the design: plasmonic absorbers have proven to be effective for realizing compact absorbers for visible and infrared radiations while metamaterials are preferably used from the terahertz to the microwaves \cite{Cui:2014bd, Landy:2008gy}. Many designs have been proposed recently, a lot of them relying on the excitation of cavity resonances of some kind\cite{Kats:2012hr, Kats:2012eb, Yao:2014ii}. All these structures being cavities, a minimum size is required that is always larger than $\lambda/20$ despite strategies to reduce the effective wavelength of the mode responsible for the resonance\cite{Moreau:2012ub, Collin:2007wp}.

Here we propose a design for a very deeply subwavelength resonant absorber whose absorption frequency can be tuned from infrared to microwave frequencies by following simple scaling laws. This absorber is universal in the sense that its optical properties are independent from the choice of the metals and dielectrics involved for its realization and it presents an almost perfect absorption for incident angles up to $30^\circ$ and for any polarization of light. The absorption mechanism is based on a negative phase shift for the light which happens at the metamaterial interfaces  and allows to built up a Fabry-Perot (FP) resonance into a near-zero dielectric thickness  of $\lambda/100$. This resonance is activated by a funneling effect through slits of few nanometers wide,  with a ratio of the period to the width of the slits that can easily be larger than 30,000.

In a first part, we concentrate on metamaterial absorber periodic in one direction and give the physical picture to explain its properties for one polarization.  The resonance responsible for the absorption is demonstrated to be a FP resonance squeezed into a sub-wavelength thickness. In a second part, we show using the previous model that the design can be tuned to absorb almost perfectly for any frequency. Finally, these results are extended to universal absorbers periodic in two directions  that present an absorption line for any polarization and a wide range of incidence angles.

\section{Theory and design of super absorbers}

The absorber consists of a deeply subwavelength grating made of nanometers slits etched in a thin metallic slab which is separated from a metallic back mirror by a dielectric spacer, Fig.1. The operating frequency range extends from the infrared to microwaves frequency when noble metals such as gold or silver are utilized (see supporting information). The upper frequency boundary is limited by the plasma frequency of the metallic medium that is typically located in the ultraviolet spectrum for noble metals but can be located in the mid-infrared for highly doped semiconductors. Without any loss of generality, we illustrate our results by considering InAsSb, a highly doped semiconductor whose plasma frequency can be tuned by playing with the doping concentration.  This material is in addition compatible with CMOS technology and  its relative permittivity is given by a Drude model $\epsilon_{InAsSb}=\epsilon_{\infty}(1-\omega_p^2/(\omega(\omega+i\gamma)))$ with $\epsilon_{\infty}=11.7$, $\omega_p= 351. 10^{12}\ rad.s^{-1}$ and $\gamma=10^{13}\ rad.s^{-1}$ \cite{NTsameGuilengui:2012jn, Taliercio:2014fn}. The spacer, of thickness $g$, is filled with a GaSb insulator (that is assumed to be non-dispersive) of refractive index $n_d=3.7$. The absorbance $A$ is deduced from the energy reflection coefficient $R$ computed using the Rigorous Coupled-Wave Analysis (RCWA)\cite{Weiss:2009kx}. First, we consider a 1D periodic set of slits of width $f=10 \ nm$ etched in the x-direction, with a pitch $d=2\ \mu m$ and a thickness $h=320\ nm$ for the metamaterial layer. Two absorption lines can be observed in normal incidence on the computed absorption spectrum, for p-polarization case ({\em i.e.} a magnetic field along the slits), Fig. 2a. The lower resonant wavelength, $\lambda_s= 11 \ \mu m$, corresponds to a cavity-like mode localized into the slits, Fig. 2b-c. Similarly to the case of Extraordinary Optical Transmission (EOT) \cite{GarciaVidal:2010ed}, a gap-plasmon having a high effective index $\bar{n}_{slit}=n_{slit}+i\kappa_{slit}$ is excited and reflected at the top and at the bottom of the metamaterial layer. The real part of the effective index is well approximated by $n_{slit}=\sqrt{1+2\delta_p/f}$  where  $\delta_p=c/\omega_p$ is the penetration depth into the metal \cite{Collin:2007wp, Pardo:2011cp}. The resonant condition for the slit reads $\lambda_s=2n_{slit} h+\lambda_{\Phi}$ where $\lambda_{\Phi}$ is a phase shift linked to the reflection coefficient of the gap-plasmon inside the metamaterial layer \cite{Moreau:2007jv, Koechlin:2013wn}. The spatial extension of the magnetic field for the second absorption line at $\lambda_r=77 \ \mu m$ indicates that it can be assimilated to a symetric Fabry-Perot resonance localized inside the spacer of sub-wavelength thickness $g=850 \ nm$ (about $\lambda/100$) well below the common quarter-wavelength criterium, Fig. 2d-e.
\begin{figure}[htbp]
\centerline{\includegraphics[width=0.8\columnwidth]{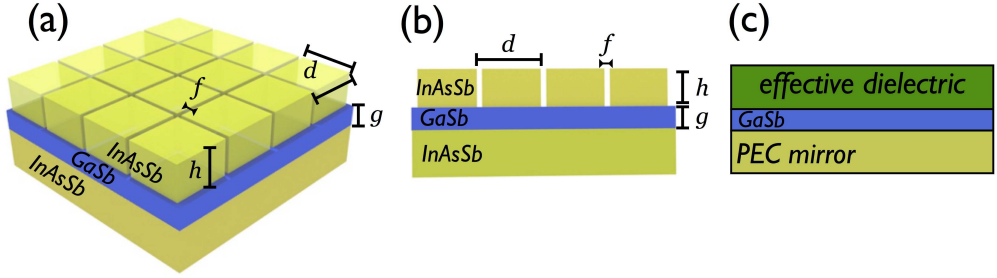}}
\caption{(a) and (b) represent respectively the 2D and 1D metamaterial absorbers made of a grating of thin slits (width $f$) periodically etched (pitch $d$) in InAsSb, a GaSb spacer and a mirror. (c) schematic of the equivalent systems consisting of an dielectric layer of effective index $\bar{n}$ and thickness $\bar{t}$, the GaSb spacer backed with a perfect electric conductor (PEC) mirror.}
\end{figure}

\begin{figure}[htbp]
\centerline{\includegraphics[width=0.8\columnwidth]{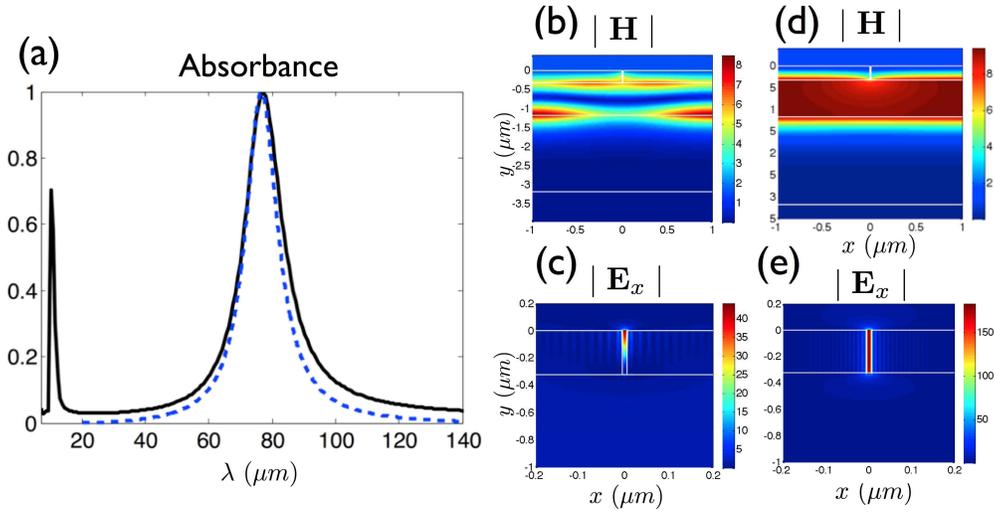}}
\caption{(a)  Absorbance for a 1D metamaterial absorber ($f=10 \ nm$, $d=2\ \mu m$, g=850 \ nm). The solid and dashed curves are respectively obtained with the exact electromagnetic simulation and with the equivalent dielectric model. (b) and (c) Maps of the modulus of the magnetic and electric fields corresponding to $\lambda_s= 11 \ \mu m$. (d) and (e) Maps of the modulus of the magnetic and electric fields corresponding to de FP resonance $\lambda_r= 77 \ \mu m$.}
\label{fig:fig2}
\end{figure}

 To explain this actual reduction of the FP cavity size, we derive an equivalent dielectric model assuming that the metallic grating is equivalent to a dielectric slab of complex refractive index $\bar{n}=n+i\kappa$ and thickness $\bar{t}$, Fig.1c. The optical property of such artificial dielectric layer is known to depend on the geometrical parameters of the grating and on the effective index of the gap plasmon by $\bar{n}=\bar{n}_{slit}d/(f+2\delta_p)$ \cite{Koechlin:2013wn}. The resonant wavelength of the gap plasmon mode is equivalently linked to the effective index and thickness by $\lambda_s=2\bar{n}\bar{t}$. With these definitions, the whole 1D absorber can be replaced by a much simpler equivalent system made of a dielectric spacer sandwiched between a back mirror assumed to be a perfect electric conductor (PEC)  and an absorbing layer of complex index $\bar{n}$ corresponding to the grating layer, Fig. 1c. As seen on Fig. 2a, this equivalent dielectric system catches the global picture of the actual metamaterial since the FP absorption line centered at $\lambda_r= 77 \ \mu m$ is well reproduced.   We try now to extract from this simpler structure, and analytical expression for the resonant wavelength, based on the different geometrical and physical parameters of the structure, which will allow us to better understand why the resonance can occur at arbitrarily large wavelengths and give us very simple design recipes for our metamaterial absorber. By taking into account the PEC back mirror, the magnetic field inside the spacer reads $H_z=B cos(k_0n_dy)$, with $k_0=2\pi/\lambda$. The electromagnetic continuity conditions applied at the interfaces lead to link by a \textbf{T} matrix the amplitude $B$ of the spacer mode to the amplitudes $I$ and $R$ of the incident and reflected waves: 
\begin{equation}
\begin{pmatrix} I  \\ R \end{pmatrix}=\textbf{T} \begin{pmatrix}  Bcos(k_0n_dg)  \\ B\frac{k_0}{in_d}sin(k_0n_dg)  \end{pmatrix}
\end{equation}
from which the amplitude of the  FP mode is expressed in a conventional formulation:
\begin{equation}
B=\frac{2\tau_{eq}}{1-\Gamma_{eq} e^{2ik_0n_dg}}I
\label{eq:mode}
\end{equation}
Here $\Gamma_{eq}=(1-\bar{n}_{eq})/(1+\bar{n}_{eq})$ is an equivalent reflection Fresnel coefficient determined by the equivalent index $\bar{n}_{eq}=\frac{n_d}{k_0}\frac{t_{1,1}}{t_{1,2}}$ related to the elements $t_{i,j}$ of the T-matrix. Remark that $\bar{n}_{eq}$ simply reduces to $n_d$ when the grating is removed leading for Eq. \eqref{eq:mode} to the case of a single dielectric slab on top of a PEC mirror. The analytical expressions for $t_{1,2}$ and $t_{1,1}$ allows to write the equivalent index as 
\begin{equation}
\bar{n}_{eq}=\frac{n_d}{\bar{n}} f(\lambda)
\label{eq:exact} 
\end{equation}
where $f(\lambda)=(1+\rho e^{2i\pi\lambda_s/\lambda})/(1-\rho e^{2i\pi\lambda_s/\lambda})$ and $\rho=(\bar{n}-1)/(\bar{n}+1)$ designates the Fresnel coefficient at the air-dielectric interface for p-polarized light. For thin slits ($d \gg f$) $\rho \simeq 1$  since $\bar{n}$ takes very high values. Thus, in the long wavelength limit, when $\lambda \gg \lambda_s$, the function $f(\lambda)$ can be approximated by $i\lambda/(\pi\lambda_s)$. These simplifications leads to write the equivalent index in the following form:  
\begin{equation}
\bar{n}_{eq}=\frac{n_d}{n} \frac{\lambda}{\pi\lambda_s}(\frac{\kappa}{n}+i)
\label{eq:neff}
\end{equation}
Introducing the figure of merit (FOM) $\mathcal{F}= n / \kappa$, the equivalent extinction coefficient $\kappa_{eq}=\frac{n_d}{n} \frac{\lambda}{\pi\lambda_s}$ is inversely linked to the equivalent refractive index by  $\kappa_{eq}=\mathcal{F}n_{eq}$. The good agreement between the exact expression of the complex equivalent index of Eq.\eqref{eq:exact} and that of the analytical one of Eq.\eqref{eq:neff} is shown in Fig. 3a. Equipped with this complex equivalent index, two optical conditions (one for phase and a second for the modulus)  can be extracted from the magnetic FP resonance condition:
\begin{equation}
1-\Gamma_{eq} e^{2ik_0n_dg}=0
\label{eq:reson}
\end{equation}
The first condition, $|\Gamma_{eq}| =1$, is satisfied for the trivial solution $n_{eq}=0$ whatever the value of the equivalent absorption $\kappa_{eq}$. In practice, almost perfect absorption higher than $98\%$ is achieved  when a near-zero equivalent index condition is satisfied, which is the case when $n_{eq}=0.1$ for instance, see Fig.3a. From the definition of $n_{eq}$, we directly derive an analytical expression for the resonant wavelength:
\begin{equation}
\lambda_r=n_{eq}\lambda_s \pi   \frac{\mathcal{F}}{n_d} n_{slit}\frac{d}{(f+2\delta_p)}.
\label{eq:lam_r}
\end{equation}
By considering the phase condition which implies that the FP mode is built up inside the spacer when the total phase is cancelled out, we obtain:
\begin{equation}
\arg(\Gamma_{eq})+2k_dg=0 \ [2\pi]
\label{eq:phase}
\end{equation}
The first term of Eq.\eqref{eq:phase} can be written in terms of the equivalent refractive index and extinction coefficient as $\arg(\Gamma_{eq})=-2\kappa_{eq}/(1-n^2_{eq})$. As seen on Fig. 3b, it is well approximated by  $\arg(\Gamma_{eff})=-2\kappa_{eq}$ when the near-zero equivalent index condition is satisfied. The ratio $\eta=g/\lambda_r$ of the spacer's thickness over the wavelength thus given by the following equation: 
\begin{equation}
\eta= n_{eq}\frac{\mathcal{F}}{2\pi n_d}  
\label{eq:gap}
\end{equation}
In order to give rules of thumb for designing the metamaterial absorbers, we set $n_{eq}=0.1$. We have found that in the long wavelength limit and whatever the metal considered, the figure of merit reaches $\mathcal{F}_{\infty}= 8/\pi$ (see supporting information). With these parameters we arrive to $\eta=1.1/100$. This shows that the FP resonance is better excited when the dielectric layer playing the role of a cavity has a thickness that is roughly only one hundredth of a wavelength. This property originates from the negative phase, $-2\kappa_{eq}$, acquire by the electromagnetic waves when they are reflected by the equivalent high index dielectric layer. It turns out the RCWA allows to access this phase rigorously, by retrieving the actual reflection coefficient on the metamaterial layer (see Supporting Information), and that such a computation totally confirms the analytical results. Inserting this phase condition into Eq.\eqref{eq:lam_r} and for nanometer slits ($2\delta \gg f$), we get a simple expression for the resonant wavelength of the FP mode:
\begin{equation}
\lambda_r= \Lambda\frac{d}{\sqrt{f}} 
\label{eq:lam_r2}
\end{equation}
with $\Lambda=2\eta\pi^2\lambda_s/\sqrt{2\delta_p}$ that can be evaluated for 10-nanometer slits to $\Lambda=3.38$. Finally, Eq.\eqref{eq:gap} and Eq.\eqref{eq:lam_r2} provide simple scaling laws for designing universal absorbers operating at arbitrary large wavelengths.

\begin{figure}[htbp]
\centerline{\includegraphics[width=0.8\columnwidth]{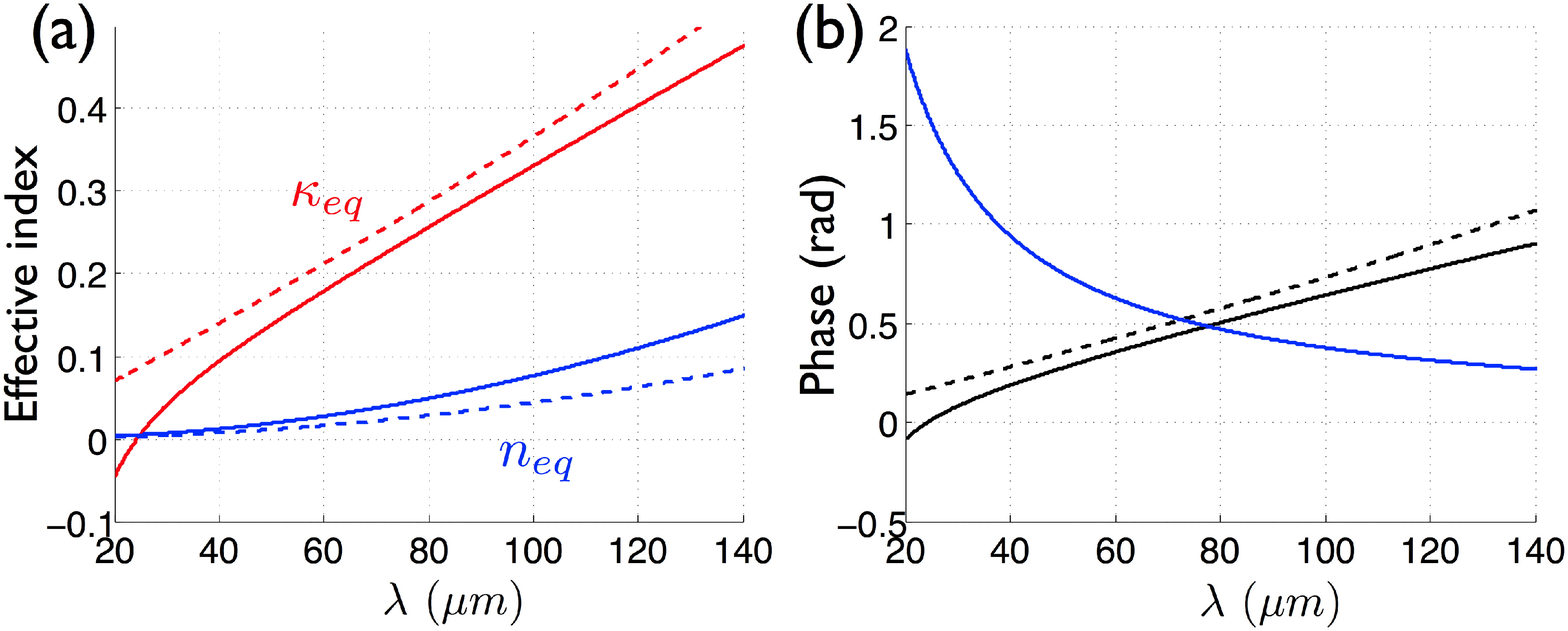}}
\caption{(a) Equivalent refractive index and extinction coefficient as a function of the wavelength obtained with Eq.\eqref{eq:exact} in solid curves and with the approximate formulation Eq.\eqref{eq:neff}. (b) Phase terms Eq.\eqref{eq:phase}: spacer phase $2k_dg$ (blue curve), $-\arg(\Gamma_{eq})$ (black curve) and $2\kappa_{eq}$ (dashed curve).}
\label{fig:fig3}
\end{figure}

\section{Perfect absorbers from infrared to microwave}

These theoretical results are confirmed by the exact electromagnetic calculations of the absorbance $A$ deduced from the reflectivity and performed in normal incidence with the RCWA method where 100 Fourier modes are used for the largest pitches. On Figure 4a, the resonant wavelengths of the FP resonance are shown for grating periods $d$ ranging from $1 \ \mu m$ to $400 \ \mu m$. In agreement with our model, the resonant wavelength is seen to be linearly linked to the period by $\lambda_r\simeq 33d$ (for $d$ in microns). The use of Eq.\eqref{eq:lam_r2} for slits of $10 \ nm$ wide leads to a slope of 33.8, thus confirming the excellent accuracy of our analytical model. This means that the structure is able to absorb microwaves with a wavelength that is more than 6 orders of magnitude larger than the slits' width (see Fig. 4a). The electric field associated to this FP magnetic resonance is actually squeezed in slits that are a million times smaller than the wavelength. This is where the absorption takes place, since the dielectric spacer is considered lossless.
\begin{figure}[htbp]
\centerline{\includegraphics[width=0.8\columnwidth]{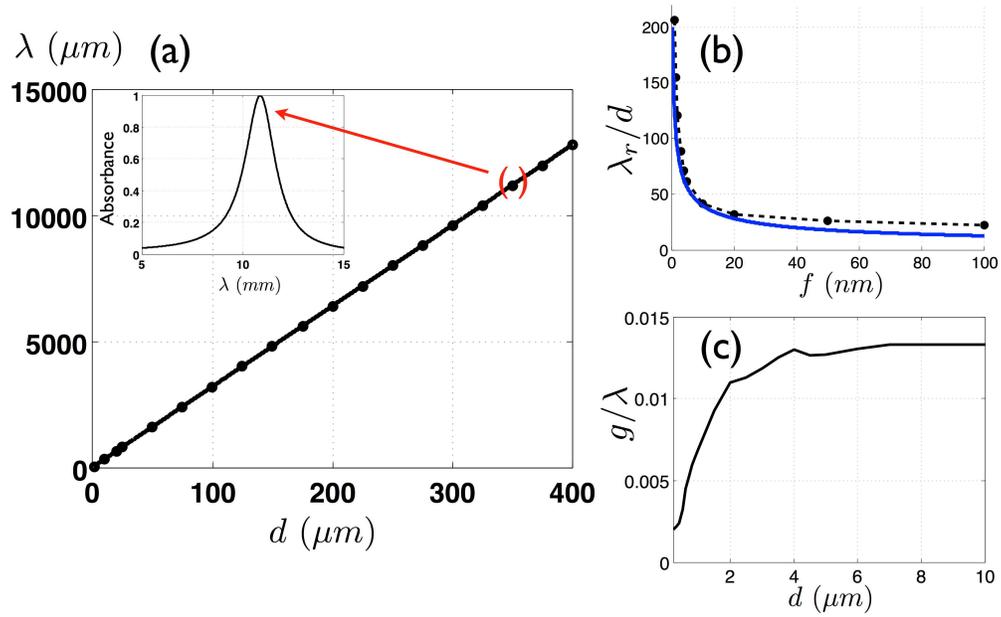}}
\caption{(a) Resonant wavelength $\lambda_r$ as a function to the pitch $d$ for a 1D absorber. The bold line corresponds to the theoretical scaling law $\lambda_r=33d$ and the dots to the exact electromagnetic computations. The inset represents the absorbance spectrum with a total absorption in the microwave range for $\lambda_r=11.2 \ mm$ when $d=350 \ \mu m$. (b) Ratio $\lambda_r/d$ with respect to the slit width, the dashed and solid curves are respectively obtained with the RCWA simulations and with Eq.\eqref{eq:lam_r2} for $d=1 \ \mu m$.  (c) Ratio of the spacer over the wavelength with respect to the pitch. $g/\lambda$ remains constant for a period larger than $8 \ \mu m$.}
\label{fig:fig4}
\end{figure}
As seen on Fig. 4b, the dimension of the slits have an important impact on the spectral position of the magnetic FP resonance. This is especially true when they are a few nanometers wide, as this dimension has a very large influence on the effective index of the gap-plasmon propagating in the slits. A quite good agreement is observed with the exact results obtained using the RCWA simulations and Eq.\eqref{eq:lam_r2} for $d=1 \ \mu m$. Beyond a pitch of $4 \ \mu m$ or equivalently for wavelengths larger than $100 \ \mu m$, the thickness of the spacer remains constant  about $g/\lambda=1.3/100$, close to the theoretical limit $\eta=1.1/100$, Fig. 4c. For arbitrarily large wavelengths,  the slits  operate as  antennas that funnel the incident waves into the spacer that constitutes the resonant cavity. The funneling factor, the ratio of the pitch to the slid width, is huge: it can be as large as 40,000 which is way above what usually happens for EOT when the resonance is located in the slits. This mechanism holds from the infrared to the microwave range despite the dispersive behavior of InAsSb and can be obtained for other metals such as silver (see supporting information).
From the application point of view, realizing  absorbers insensitive to the incident angle and to the polarization of light is a crucial issue. We address these problems by considering 2D metamaterial absorbers made of a square array of thin slits (width $f=10 \ nm$) separated by a pitch $d=2 \mu m$, Fig. 1a. We illustrate these properties for a targeted absorption line at $\lambda_r=70 \mu m$ leading to a spacer's thickness  $g=850 \ nm$, a pitch  $d=2 \ \mu m$ and a grating's thickness $h_r=320 \ nm$. More than $90 \%$ of the incident radiation is absorbed by the metamaterial for incident angles up to $50^{\circ}$ and the absorbance reaches $70 \%$ at grazing incidence for $70^{\circ}$, Fig. 5a. The efficiency of the absorber is also seen to be insensitive to the polarization of light: the structure can simply be seen as two crossed gratings, each one being responsive to one polarization only. In normal incidence, the absorbance thus remains constant whatever the polarization in normal incidence.

\begin{figure}[htbp]
\centerline{\includegraphics[width=0.8\columnwidth]{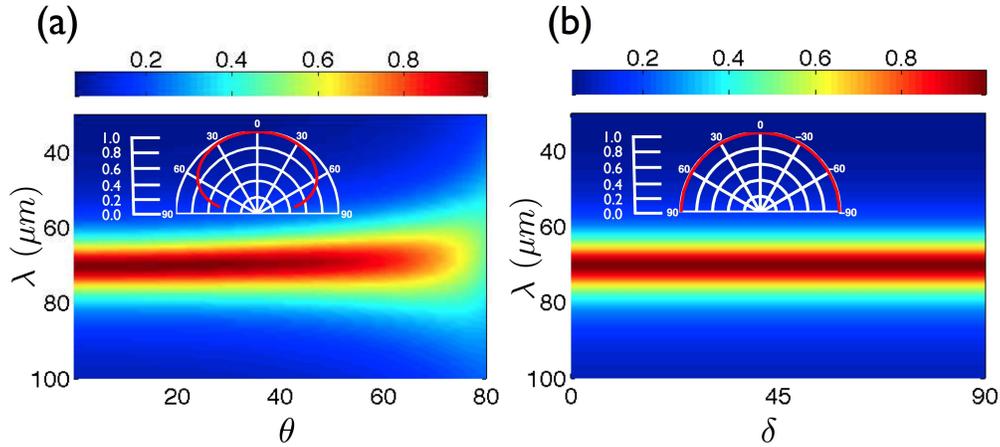}}
\caption{(a) Absorbance with respect to the incident angle $\theta$ and for a polarization angle $\delta=0^{\circ}$.  The inset represents the polar plot of the absorbance computed for the absorption line $\lambda_r=70 \ \mu m$.  (b) Absorbance with respect to  the polarization angle $\delta$ and for normal incidence (TM and TE polarizations cases are respectively defined by $\delta=0^{\circ}$ and $\delta=90^{\circ}$. In the inset shows the polar plot of the absorption line as a function of  $\delta$.}
\label{fig:fig5}
\end{figure}

%%%%%%%%%%%%%%%%%%%%%%%%%%%%%%%%%%%%%%%%%%%%%%%%%%%%%%%%%%%%%%%%%%%%%
\section{Conclusion}

We have proposed a metamaterial resonant absorber whose absorption line can be chosen in any frequency range, from optics to microwaves, by following simple scaling laws, essentially.  Our approach allows to design a perfect absorber that working for  any frequency, independently of the materials that are considered. Almost perfect absorption can be obtained whatever the polarization over a broad incident angle range. This is why we think our design can be said to be {\em universal}. The metamaterial layer controlling the response of the structure allows to reduce to $\lambda/100$ the thickness of the spacing layer constituting a resonant cavity on which the device is based. The absorption takes place in slits that are no more than a few nanometers wide, despite wavelengths that are 1000,000 times larger. All the incoming radiation in funneled through these slits despite a ratio of 1 to 40000 between the slit width and the period. We have derived an analytical model thoroughly describing the electromagnetic response of the device that proved very accurate despite these extreme and unprecedented ratios. Universal absorbers constitute ultra-thin and flexible solutions for absorbing any kind of electromagnetic radiation especially at terahertz frequencies where absorbers, sensors and emitters are usually difficult to design. Integrating electric contacts into the device or making it dynamically tunable is something that can totally be envisaged, broadening even more the potential of our design.

\begin{acknowledgement}
This work is supported by the Agence Nationale de la Recherche of France, project ``Physics of Gap-Plasmons'' (ANR-13-JS10-0003).
\end{acknowledgement}

\bibliographystyle{plain}


\end{document}